\documentstyle[pra,aps,graphicx]{revtex}
\begin{document}
\draft
\twocolumn[\hsize\textwidth\columnwidth\hsize\csname 
@twocolumnfalse\endcsname
\title{Control of conditional pattern with polarization entanglement} 
\author{D. P. Caetano$^{*,1}$, C. H. Monken$^{2}$, and  P.H. Souto Ribeiro$^{1}$}
\address{$^{1}$Instituto de F\'{\i}sica, Universidade Federal do Rio de 
Janeiro, Caixa Postal 68528, Rio de Janeiro, RJ 21941-972, Brazil}
\address{$^{2}$Departamento de F\'{\i}sica, 
Universidade Federal de Minas Gerais, Caixa Postal 702, 
Belo Horizonte, MG 30123-970, Brazil} 
\date{\today}
\maketitle
\begin{abstract}
Conditional interference patterns can be obtained with twin photons from spontaneous parametric down-conversion and the phase of the pattern can be controlled by the relative transverse position of the signal and idler detectors. Using a configuration that produces entangled photons in both polarization and transverse momentum we report on the control of the conditional patterns by acting on the polarization degree of freedom.
\end{abstract}

]

\section{Introduction}

Spatial optical patterns have been subject of interest along the development of classical and quantum optics. For the case of the light produced in parametric down-conversion, spatial patterns are present in tranverse correlation functions\cite{1,2} and distribution of quantum fluctuations\cite{3}. These spatial patterns have been demonstrated to carry information about coherence properties\cite{4}, mode structure\cite{5}, and orbital angular momentum\cite{6}, for example. 

Most of the observations of quantum effects in optics rely on conditional measurements\cite{7}. Conditional interference patterns in the fourth-order correlation function have been observed with twin photons produced by spontaneous parametric down-conversion in several experiments\cite{2,8,9,10,11,12,13}. For the case of spatial patterns, diffracting masks placed along the propagation of the beams are very often used\cite{2,9,10,11}. The conditionality is observed by varying the relative transverse position between signal and idler detectors. Here we focus on the configuration described in Ref.\cite{13} that allows generating spatial patterns without diffracting masks and to prepare a state which is simultaneously entangled in polarization and transverse momentum with controlled degree of entanglement. The entanglement in transverse momentum leads to a quantum interference, that can be revealed through conditional spatial patterns\cite{12}. This interference pattern have been demonstrated to be useful to measure the degree of entanglement in polarization\cite{13} and to manipulate the entanglement in transverse momentum through the polarization \cite{14}. In this work we demonstrate experimentally the control of the conditional spatial pattern by acting on the polarization degree of freedom.

\section{Theory}

\begin{figure}[h]
%\vspace*{4cm}
%\special{wmf:fig1.wmf x=7cm y=4cm}
\includegraphics*[width=7cm]{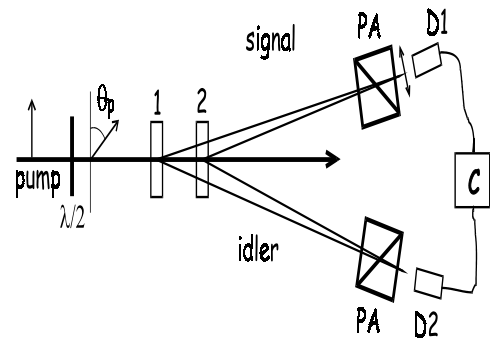}
\caption{Experimental setup.}
\label{fig1}
\end{figure}

Consider the situation sketched in Fig.\ref{fig1}. 
Two non-linear crystals are pumped by the same laser beam and at the detection plane, signal and idler beams originating at different crystals overlap. The laser beam is 45 degrees polarized to pump equally both crystals. One of the crystals produce twin photons linearly
polarized in the horizontal direction(H) and the other vertically polarized(V) twin photons.
The strength of the pump interaction with each crystal can be controlled by a halfwave plate acting on the laser beam polarization.
Before detectors, polarization analyzers projects all beams onto 45 degrees linear polarization. 
The signal beam passes through a halfwave plate before the analyzer. 
Therefore, just before detection, the state of the field can be written as:

\begin{eqnarray}
\label{eq1}
|\psi\rangle = \alpha \cos (\frac{\pi}{4} - 2\theta) |1_{H}\rangle_{i1} |1_{H}\rangle_{s1}
|0\rangle_{i2}|0\rangle_{s2} + \\ \nonumber
\beta \mbox{e}^{i\phi}\sin (\frac{\pi}{4} - 2\theta) 
|0\rangle_{i1} |0\rangle_{s1}|1_{V}\rangle_{i2}|1_{V}\rangle_{s2},
\end{eqnarray}
where $\alpha$ and $\beta$ are coefficients taking into account the strength of the
pump beam in each polarization direction (H and V) and normalization. $\theta$ is the
angle between the polarization of the signal beam after the halfwave plate and the horizontal direction. $\mbox{e}^{i\phi}$
is the phase difference between the two states and i(s)$_{1(2)}$ refer to 
signal(s) and idler(i) beams originated in crystals 1 and 2. The phase difference
associated to the different possible paths for the beams can be better understood
considering them coming from the field propagation, as in Ref.\cite{12},
but it can also be considered in the state as well. This phase difference gives rise to the conditional interference patterns of our interest. 

The coincidence counting rate is given by:

\begin{eqnarray}
\label{eq2}
C=\alpha^2 \cos^2(\pi/4 - 2\theta) + \beta^2 \sin^2 (\pi/4 - 2\theta) +\\ \nonumber
(\alpha \beta/2) \sin 2(\pi/4 - 2\theta) \cos \phi.
\end{eqnarray}
As we can see, the coincidence rate has an oscillating behavior depending on the polarization angle of the signal beam $\theta$ and on the phase difference $\phi$.

From the above it is seen that actions on the halfwave plate(in the signal beam) 
affects the degree of entanglement and therefore the visibility of the patterns. However, even for
maximally entangled states, it is possible to change the phase of the interference pattern,
by choosing an angle $\theta$ which changes the sign of the $sin$ function in the
last term of Eq.\ref{eq2}. 

\section{Experiment}
The experiment have been performed with two 1cm long LiIO$_{3}$(Lithium Iodate) nonlinear crystals cut for type-I phase matching pumped by a He-Cd laser operating at 442nm as depicted in Fig.\ref{fig1}. The separation between the crystals is about 1cm. The polarization of the pump beam is turned into 45 degrees from the horizontal direction
by a halfwave plate before the crystals. The optical axes of the crystals are aligned so that the first crystal is pumped by the horizontally polarized component and the second crystal is pumped by the vertically polarized component. Signal and idler
photons at 884nm are detected before passing through polarizing beam splitters, interference filters with 10nm bandwidth and entrance slits. Before the signal polarizing beam splitter, a halfwave plate is used for rotating the input polarization. The detectors are avalanche photodiode photon counting modules placed about 1m from the crystals.

\begin{figure}[h]
%\vspace*{6cm}
%\special{wmf:fig2.wmf x=8cm y=6cm}
\includegraphics*[width=8cm]{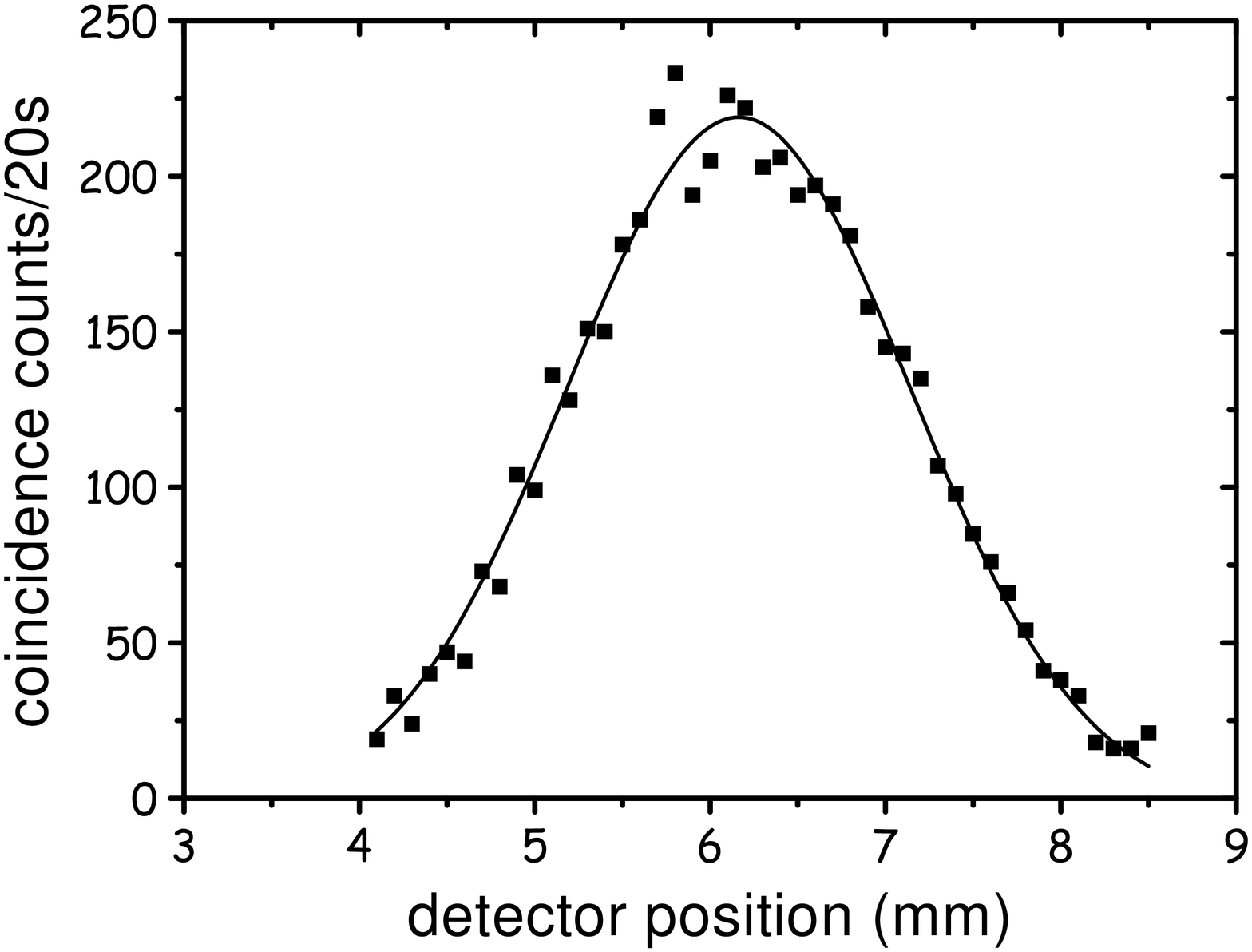}
\caption{Crystal 1 coincidence profile scanning D1 with D2 fixed. Full line is a guassian fit.}
\label{fig2}
\end{figure}

\begin{figure}[h]
%\vspace*{6cm}
%\special{wmf:fig3.wmf x=8cm y=6cm}
\includegraphics*[width=8cm]{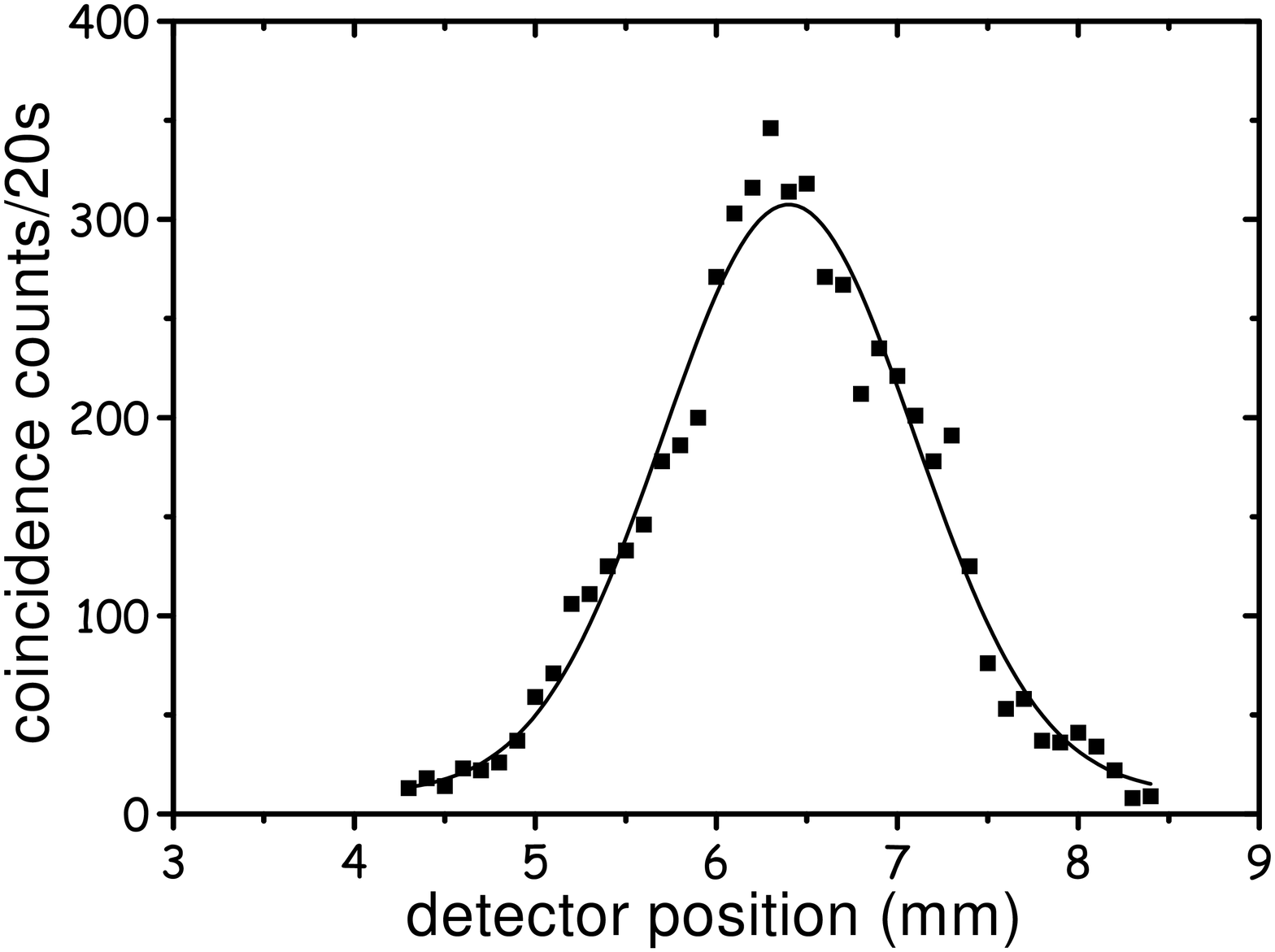}
\caption{Crystal 2 coincidence profile scanning D1 with D2 fixed. Full line is a Gaussian fit.}
\label{fig3}
\end{figure}

We have measured the coincidence
profile for the photons originating in each crystal. For doing so, the pump beam polarization
is first oriented vertically, so that it only pumps the first crystal. The signal detector is
horizontally scanned while the idler detector is kept fixed. The polarization analyzers(waveplate and polarizing beam splitters) were both oriented to 45 degrees transmission. The scan is performed with a 0.2mm slit in the detector entrance. The result is shown in Fig.\ref{fig2}. The procedure is repeated when the pump beam polarization is oriented horizontally, in order to only pump the second crystal. The
result is shown in Fig.\ref{fig3}. Gaussian fittings to these curves indicate the position of
the peaks at x$_{c}$=6.2mm and x$_{c}$=6.4 mm, respectively. This shows that the twin
photons from both crystals are correlated around the position x$_{c}$=6.3mm of the
signal detector when the idler detector is fixed.

\begin{figure}[h]
%\vspace*{6cm}
%\special{wmf:fig4.wmf x=8cm y=6cm}
\includegraphics*[width=8cm]{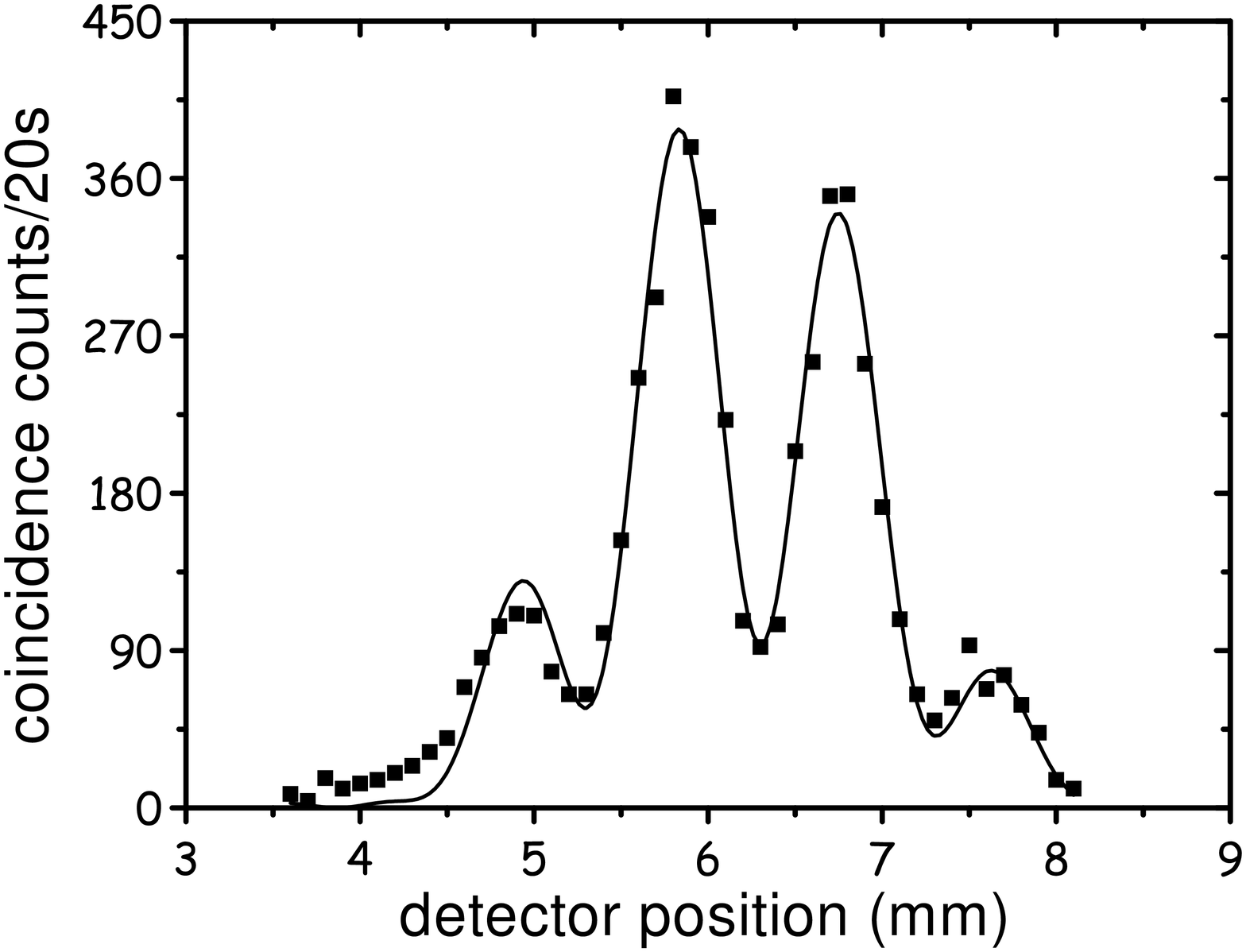}
\caption{Coincidence profile for both crystals. D1 is scanned horizontally and the signal waveplate is set to $45$ degrees. D1 fixed. Full line is a nonlinear fit.}
\label{fig4}
\end{figure}

Now, the pump beam polarization is set to 45 degrees, so that it pumps both crystals
with the same strength. Performing the horizontal scan with signal detector as before,
and leaving both polarization analyzers oriented to 45 degrees, we obtain the
spatial pattern shown in Fig.\ref{fig4}. At the position around x$_{c}$=6.3mm the coincidence
counting rate is much lower than that for twin photons originating in each crystal individually.

\begin{figure}[h]
%\vspace*{6cm}
%\special{wmf:fig5.wmf x=8cm y=6cm}
\includegraphics*[width=8cm]{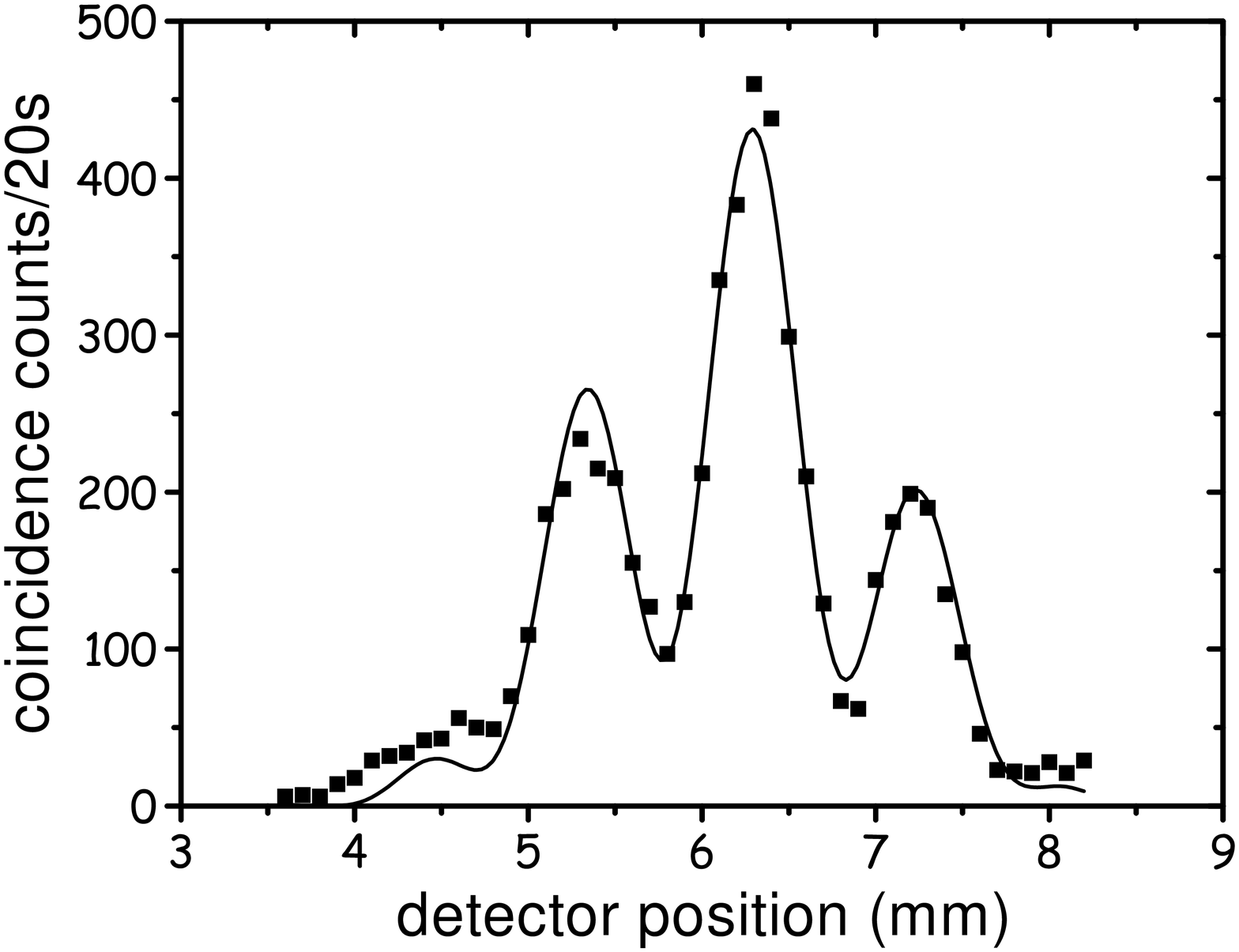}
\caption{Coincidence profile for both crystals. D1 is scanned horizontally and the signal waveplate is set to zero degrees. D1 fixed. Full line is a nonlinear fit.}
\label{fig5}
\end{figure}

The second step is to demonstrate that acting on the polarization, the pattern can be changed. For doing so the waveplate in the
signal beam is turned 45 degrees and the input state arriving at the polarizing beam splitters
is changed from:

\begin{equation}
\label{eq3}
|\psi\rangle = \alpha |1_{H}\rangle_{i1} |1_{H}\rangle_{s1}
|0\rangle_{i2}|0\rangle_{s2} + 
\beta |0\rangle_{i1} |0\rangle_{s1}|1_{V}\rangle_{i2}|1_{V}\rangle_{s2}
\end{equation}
to
\begin{equation}
\label{eq4}
|\psi\rangle = \alpha |1_{H}\rangle_{i1} |1_{V}\rangle_{s1}
|0\rangle_{i2}|0\rangle_{s2} + 
\beta |0\rangle_{i1} |0\rangle_{s1}|1_{V}\rangle_{i2}|1_{H}\rangle_{s2}.
\end{equation}
Fig.\ref{fig5} shows the coincidence profile scanning signal detector and keeping idler detector fixed in the same position of the previous measurement. 
Now at the position around x$_{c}$=6.3 mm there is a coincidence peak
instead of a valley. Using the usual double-slit diffraction function to fit the experimental data we found the visibilities of the patterns showed in Fig.\ref{fig4} and \ref{fig5} to be 64\% and 62\%, respectively.

\begin{figure}[h]
%\vspace*{6cm}
%\special{wmf:fig6.wmf x=8cm y=6cm}
\includegraphics*[width=8cm]{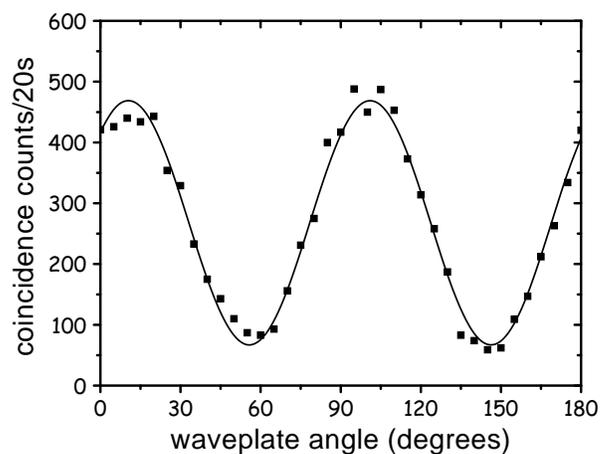}
\caption{Coincidence counts for both crystals. D1 and D2 are fixed while the signal waveplate is turned. Full line is a nonlinear fit.}
\label{fig6}
\end{figure}

To show that the polarization entanglement is present together with transverse momentum a Bell type measurement have been performed. Fig.\ref{fig6} shows interference fringes scanning the halfwave plate angle and signal detector fixed at x$_{c}$=6.3 mm presenting 75\% visibility. This measurement is enough to demonstrate entanglement without perform a complete Bell measurement and show that some Bell inequality is violated. This kind of source of polarization entangled states have been demonstrated to produce all family of Bell states with good quality in a configuration with 1mm separated two thin crystals\cite{15}.

\section{Discussion}

The interference patterns presented in Figs.\ref{fig4} and \ref{fig5} show that
the phase of the conditional interference pattern can be controlled by
changing the input polarization state at the analyzers. This is a consequence
of the coupling between polarization and transverse degrees of freedom of the 
twin photons. As the interference is conditioned to the detection at the idler side, the phase, visibility and even the length of the oscillations obtained by the displacement of the signal detector,
depend on the idler detection properties\cite{12}. In the case presented above,
the idler detection properties were kept fixed, so that only the polarization
state was changed. 

In order to observe this effect, the presence of the polarization analyzers,
is required before detectors. This means that the polarization gives us
control over the spatial correlation between twin photons by post-selection. 
In our measurements we were able to pass from anti-correlation to correlation
in the position corresponding to the individual(i.e. for each crystal
independently) coincidence peaks. 

In fact, for the usual double-slit experiment with single photons and intensity
measurements, it is possible to have a similar control over the phase of the
pattern. For doing that, we could place waveplates in each one of the slits
and perform the measurement of the intensity pattern after some linear
polarization analyzer. The principle for acting on the phase of the pattern
is the same, but the consequences are different. When we act on the phase of
a conditional pattern, we are actually changing the spatial correlation
between photons. For instance, a collinear version of our set-up with signal and idler having transverse coordinates with opposite sense, the change in the phase of the pattern would 
imply in the switching from bunching to anti-bunching spatial statistics,
in the same fashion as it was done in Ref.\cite{16}.

\section{Conclusion}

In conclusion, we have demonstrated experimentally how to control the
phase of a conditional interference pattern, manipulating the polarization
degrees of freedom of a state of the electromagnetic field entangled in transverse momentum and polarization. This kind of manipulation opens possibility to control properties of the field state in a given degree of freedom conditioned to the action on another one. It also suggests that it is possible to prepare  special states of the light with controlled statistics.

\begin{acknowledgments}
Financial support was provided by Brazilian agencies CNPq, PRONEX, FAPERJ, 
FUJB and Institutos do Mil\^enio-Informa\c c\~ao Qu\^antica.
\end{acknowledgments}

\end{document}